\documentstyle[epsfig]{aipproc}

\newcommand{\kpnn}    {\mbox{$K^+ \! \rightarrow \! \pi^+ \nu \overline{\nu}$ }}
\newcommand{\klpnn}   {\mbox{$K^\circ_L \! \rightarrow \! \pi^\circ \nu \overline{\nu}$ }}
\newcommand{\kzpnn}    {\mbox{$K \! \rightarrow \! \pi \nu \overline{\nu}$ }}
\newcommand{\kzpen}    {\mbox{$K\!\rightarrow\!\pi e \nu_e$ }}
\newcommand{\kpp}     {\mbox{$K^+ \! \rightarrow \! \pi^+ \pi^\circ$ }}
\newcommand{\kmn}     {\mbox{$K^+ \! \rightarrow \! \mu^+ \nu_\mu$ }}
\newcommand{\vtd}     {\mbox{$V_{td}$ }}
\newcommand{\Vtd}     {\mbox{$| V_{td} |$ }}
\newcommand{\bpsiks}  {\mbox{$B^\circ_d \! \rightarrow \! \psi K^\circ_S$ }}

\begin{document}
\title{
\vspace{-1.5cm}
\rightline{\small\rm BNL--67629}
\vspace{-0.3cm}
\rightline{\small\rm August 15, 2000}
\vspace{1cm}
Future Kaon Programs at BNL, FNAL
\footnote{ To be published in the {\it Proceedings of the
7$^{th}$ Conference on the Intersections of Particle and Nuclear Physics;
Quebec City, Canada, May 22-28, 2000};
Z.~Parsa and W.~Marciano, Eds.  }
 }

\author{S.H.~Kettell}
\address{Brookhaven National Laboratory}

\maketitle

\begin{abstract}
Future kaon decay programs at BNL and FNAL are discussed.  The primary
focus of these programs is the measurement of the golden modes, \klpnn
and \kpnn.  The observation of \kpnn by E787 at BNL is the first step
in a series of measurements which will completely determine the
unitarity triangle within the kaon system.

The next step after E787 in the measurement of B(\kpnn) will be the
E949 experiment at BNL that is currently under construction. This
experiment, building on the experience of E787 and making use of the
intense AGS proton beam, is scheduled to run in FY01--03 and to
observe ${\cal O}(10)$ SM events with a small and well-understood
background. The proposed CKM experiment at FNAL would take the next
step, using a decay-in-flight technique and a 22 GeV/c RF-separated kaon
beam from the Main Injector, to observe ${\cal O}(100)$ SM
events.

At the same time, two concepts for the measurement of B(\klpnn)
have been developed. One of these, building on the experience of
KTeV with a `pencil' $K_L$ beam, has been proposed at FNAL as KAMI.
The other, with a measurement of the kaon momentum in a large angle
$K_L$ beam derived from a bunched proton beam, has been proposed at
BNL as KOPIO.
\end{abstract}

\section*{Introduction}

The decays \klpnn and \kpnn are two of the `golden modes' for
measuring CKM parameters. Measurement of the branching ratio B(\kpnn)
provides a clean and unambiguous determination of the CKM matrix
element \Vtd, in particular of the quantity $|\lambda_t|\equiv
|V^*_{ts}V_{td}|.$ Measurement of the direct-CP-violating decay \klpnn
will cleanly determine the imaginary part of $\lambda_t$, $Im(\lambda_t)$.

The theoretical uncertainty in \kpnn is quite small ($\sim$7\%) and
even smaller in \klpnn ($\sim$2\%), as the hadronic matrix element can
be extracted from the \kzpen branching ratio.  The \kzpnn branching
ratios have been calculated to next-to-leading-log
approximation~\cite{bb1}, complete with isospin violation
corrections~\cite{marciano} and two-loop-electroweak
effects~\cite{bb2}. Fits based on the best current data for the CKM
matrix elements give branching ratios of~\cite{bb3}
\begin{eqnarray}
B(\kpnn)  & = & (8.2\pm3.2)\times10^{-11} \\ \nonumber
B(\klpnn) & = & (3.1\pm1.3)\times10^{-11} .
\end{eqnarray}
These branching ratios are very small and, with two neutrinos in the
final state, both of these experiments are challenging.

\boldmath
\section*{Measurement of \Vtd from \kpnn}
\unboldmath

\begin{figure}[h] 
  \epsfig{file=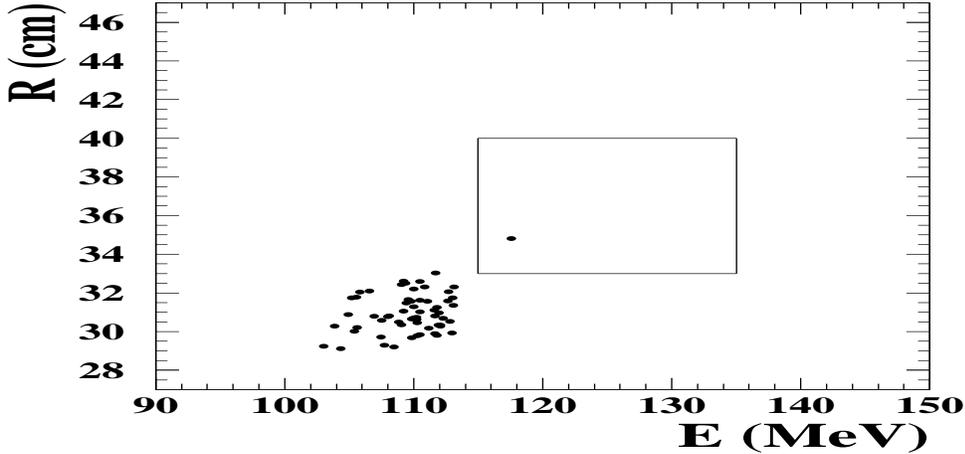,height=2.5in,width=5in}
\caption{ Range vs. Kinetic energy plot of the final sample. 
The events near $E=108$~MeV are $K_{\pi 2}$ background. The box
indicates the accepted region for \kpnn events.}
\label{fig:r_e}
\end{figure}

The E787 experiment at BNL was designed to search for \kpnn and
reported the first observation of \kpnn from analysis of the 1995 data
set~\cite{pnn95}. A new analysis of the 1995 data combined with the
1996 and 1997 data sets, has reduced the background levels by about a
factor of three.  A plot from the 1995--97 data set~\cite{e787_pnn} of
the range vs. energy of events passing all other \kpnn criteria is
shown in Figure~\ref{fig:r_e}.
One event is observed in the signal region with a measured background
of $0.08\pm0.02$ events.  The branching ratio is B(\kpnn) =
$(1.5^{+3.4}_{-1.2}) \times 10^{-10}$. From this measurement, a limit
on \Vtd of $0.002 < |\vtd| < 0.04$ can be derived, as well as the
following limits on $\lambda_t \equiv V^*_{ts}\vtd$: $| Im(\lambda_t)
| < 1.22\times10^{-3}$, $-1.10\times10^{-3} < Re(\lambda_t) <
1.39\times10^{-3}$, and $1.07\times10^{-4} < | \lambda_t | <
1.39\times10^{-3}$.  The E787 experiment has finished running and the
final sensitivity, based on the complete 1995--98 data set, should
reach the SM expectation for \kpnn.

\subsubsection*{E949 at BNL}

A new experiment under construction, E949, is expected to run in
2001--03, symbiotically with RHIC. E949 will use the AGS proton beam,
between fills of RHIC, for approximately 20 hours/day.  The E787
experiment has already demonstrated sufficient background rejection
($\sim$10\% of the SM signal) for a very precise measurement of
B(\kpnn).  Taking advantage of the very large AGS proton flux and the
experience gained with the E787 detector, E949 with modest upgrades
should observe $\cal O$(10) SM events in a two year run. The
background is small and well-understood.

\subsubsection*{CKM at FNAL}

The CKM experiment was proposed in 1998 and has been pursuing R\&D
towards a full technical proposal in 2001 as E905 at FNAL. It would
run simultaneously with the Tevatron collider using protons from the
Main Injector that are not needed for the collider and extract them
over a long spill ($\sim$1 sec).  CKM plans to collect $\cal O$(100)
SM events with a background of $\cal O$(10) events, starting sometime
after 2005. This experiment will use an intense RF-separated 22 GeV/c
kaon beam derived from the Main Injector. This novel K$^+$
decay-in-flight technique will obtain redundant kinematic measurements
from independent momentum and velocity spectrometers.  The kaon
momentum will be measured in a Si spectrometer and the pion momentum
in straw-tube drift chambers in the vacuum decay region. The
velocities of the kaon and pion will be measured in RICH counters. Two
large Pb-scintillator photon veto systems reduce backgrounds from \kpp
decays and a muon veto system reduces background from \kmn decays.

\boldmath
\section*{Measurement of $I\lowercase{m}(\vtd)$ from \klpnn}
\unboldmath

\begin{figure}[htbp]
\epsfig{file=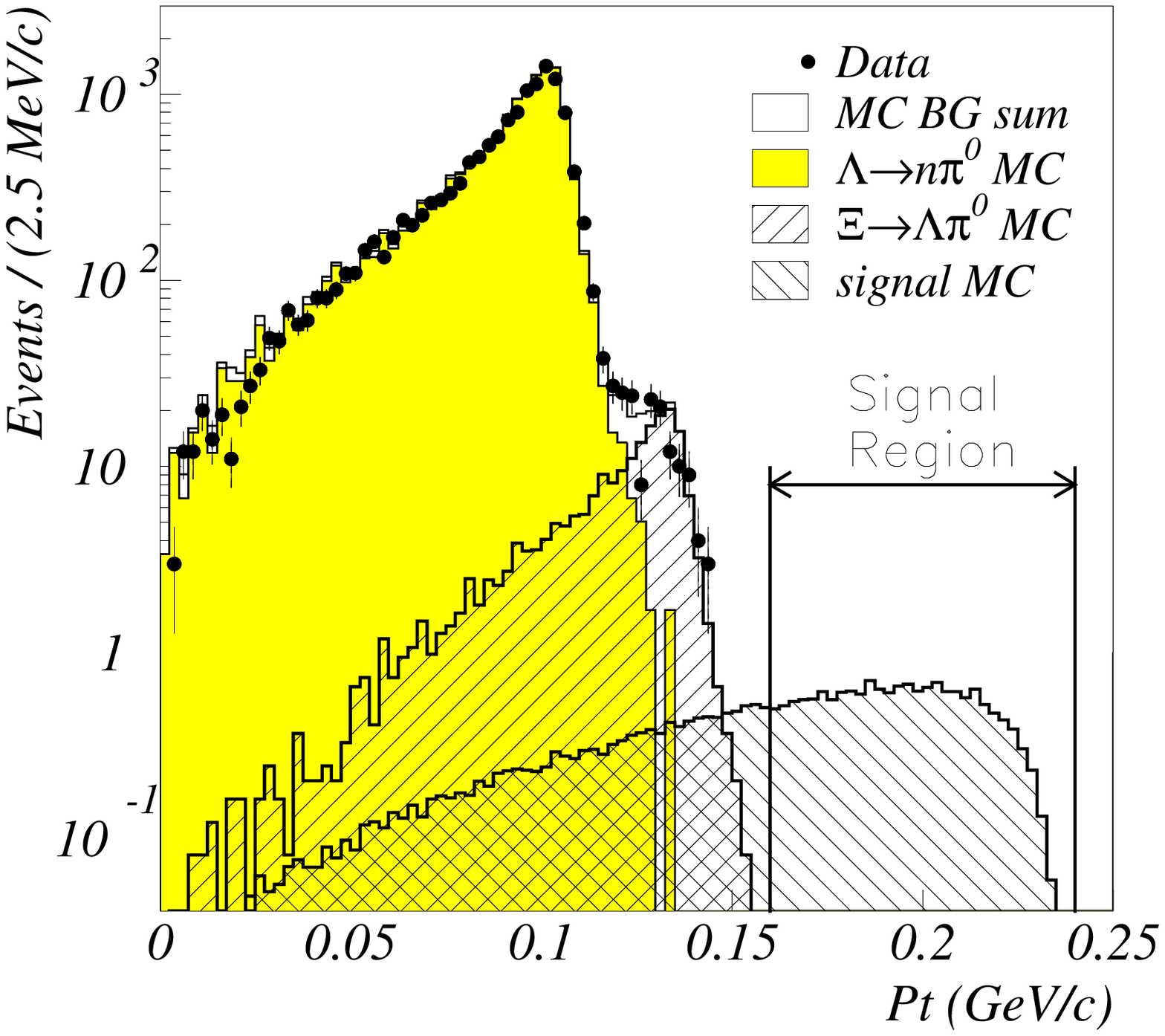,height=2.5in,width=5.in,angle=0}
\caption{Final KTeV \klpnn ($\pi^\circ\rightarrow e^+e^-\gamma$) 
data sample collected during 1996--1997 after all cuts. No
\klpnn events are seen above $P_T = 160$ MeV/$c$.
\label{fig:e799_pnn}}
\end{figure}

Presently, the best limit on \klpnn is derived in a model-independent
way~\cite{grossman} from the E787 measurement of \kpnn:
\begin{eqnarray}
B(\klpnn) & < & 4.4\times B(\kpnn) \\ \nonumber
         & < & 2.6\times10^{-9} \; \; (90\%\,{\rm CL}).
\label{eq:pnn}
\end{eqnarray}
The best direct limits come from the KTeV experiment at FNAL.  KTeV
used a narrow `pencil' beam to define the transverse vertex position of
$\pi^\circ\rightarrow\gamma\gamma$ decays in a one-day test run and
observed one background event, probably from a neutron interaction.
From this special run, a 90\%-CL limit~\cite{e799_pnn_gg} of B(\klpnn)
$< 1.6\times 10^{-6}$ was established.  A better limit is obtained
using the $\pi^\circ\rightarrow e^+e^-\gamma$ decay, which is
inherently a factor of 80 less sensitive but has the significant
advantage of a precise vertex location.  Since the vertex location is
known a larger, more intense kaon beam can be used; and the background
levels are lower as the transverse momentum is known with better
precision.  In the full 1997 KTeV data set no events were seen, and at
the 90\% confidence level, $B(\klpnn) < 5.9\times
10^{-7}~\cite{e799_pnn}.$ The $P_T$ distribution of
$\pi^\circ\rightarrow e^+e^-\gamma$ events passing all other cuts can
be seen in Figure~\ref{fig:e799_pnn}.
The expected background was $0.12^{+0.05}_{-0.04},$ mainly from
$\Lambda\rightarrow n\pi^\circ$ and
$\Xi^\circ\rightarrow\Lambda\pi^\circ.$

The next generation of \klpnn experiments, all using the
$\pi^\circ\rightarrow\gamma\gamma$ decay mode, will start with E391a at KEK,
which hopes to reach a sensitivity of $\sim$$10^{-10}$, using a 
technique similar to KTeV. Although the
reach of E391a is not sufficient
to observe a signal at the standard model level,
the experiment will be able to rule
out large enhancements from new physics
and learn more about how to do this difficult experiment.
It is designed around a pencil $K_L$ beam, a high-resolution
crystal calorimeter, and very efficient photon veto systems. This
experiment would eventually  move to the JHF and aim for a
sensitivity of ${\cal O} (10^{-14})$.
Two other major efforts to observe and measure \klpnn are
KAMI and KOPIO.

\subsubsection*{KAMI at FNAL}

The KAMI collaboration submitted an expression of interest at FNAL for
an experiement to measure B(\klpnn). KAMI, like CKM, will make use of
a slow extracted spill from the Main Injector, simultaneous with the
operation of the Tevatron. KAMI plans to reuse the excellent KTeV CsI
calorimeter, with a single high intensity pencil $K_L$ beam directed
through a hole in the middle.  The decay volume upstream of the
calorimeter will be 
surrounded by a hermetic, highly efficient array of photon veto
detectors.  An additional photon detector will catch photons escaping
along the beam. The current design of the KAMI detector includes a
fiber tracker to expand the number of secondary modes to be studied. KAMI
expects to detect ${\cal O}$(100) SM events in a couple of years of
running with a background of $\sim$40\% of the SM signal.

\subsubsection*{KOPIO at BNL}

The KOPIO experiment at BNL has been given scientific approval and is
currently undergoing funding review. It plans to run at the AGS after
the completion of E949 and in the same mode, with $\sim$20 hours per
day available between RHIC fills.  KOPIO will reconstruct the kaon
center of mass using a bunched proton beam and a very low momentum
$K_L$ beam.  This technique allows for two independent criteria to
reject background, photon veto and kinematics---allowing background
levels to be directly measured from the data---and encourages further
confidence in the signal by measuring the momentum spectrum of the
decay.  The necessary kaon flux will be obtained using the large
available AGS proton current. The low-energy beam also substantially
reduces backgrounds from neutrons and other sources. After three years
of running, 65 standard-model events are expected with a S/B $\ge$
2:1.

\section*{Conclusions}

The unprecedented sensitivities of rare kaon decay experiments and the
recent discovery of \kpnn have opened doors to the measurement of the
unitarity triangle completely within the kaon system.  Significant
progress in the determination of the fundamental CKM parameters will
come from the generation of experiments that is now starting.  These
measurements can provide critical, unambiguous determination of the
standard-model {\it CP} violation parameters. Comparison with the
B-system will then over-constrain the triangle and test the SM
explanation of {\it CP} violation:
\begin{itemize}
  \item Comparison of the angle $2\beta$ from the ratio
B(\klpnn)/B(\kpnn) and the CP asymmetry in the decay
\bpsiks will provide one of the most important tests~\cite{grossman,sinb}.
  \item Comparison of the magnitude $|\vtd|$ from \kpnn and the ratio of
the mixing frequencies of $B_s$ to $B_d$ mesons will also provide
an important test with small theoretical uncertainty~\cite{bb3}.
\end{itemize}

 
\section*{Acknowledgments}

I wish to thank many people for discussions regarding this talk:
particularly, Greg Bock, Laurie Littenberg, Ron Ray, Tony Barker, Bob
Tschirhart, Robin Appel and Peter Cooper.  This work was supported
under U.S. Department of Energy contract \#DE-AC02-98CH10886.

\def\Journal#1#2#3#4{{#1}{\bf #2}, #3 (#4)}

\def\ARNPS{{\it Ann.\ Rev.\ Nucl.\ Part.\ Sci.\ }}
\def\NCA{{\it Nuovo Cimento }}
\def\NIM{{\it Nucl.\ Instrum.\ Methods }}
\def\NIMA{{\it Nucl.\ Instrum.\ Methods} \bf A}
\def\NPB{{\it Nucl.\ Phys.} \bf B}
\def\PLB{{\it Phys.\ Lett.}  \bf B}
\def\PRL{{\it Phys.\ Rev.\ Lett.\ }}
\def\PRD{{\it Phys.\ Rev.} \bf D}


\begin{references}
\bibitem{bb1} Bucahalla~G. and Buras~A., \Journal{\NPB}{412}{106}{1994}.
\bibitem{marciano} Marciano~W.J. and Parsa~Z., \Journal{\PRD}{53}{R1}{1996}.
\bibitem{bb2} Bucahalla~G. and Buras~A., \Journal{\PRD}{57}{216}{1998}.
\bibitem{bb3} Bucahalla~G. and Buras~A., \Journal{\NPB}{548}{309}{1999}.\label{ref:bb3}
\bibitem{pnn95} Adler~S., {\it et al.}, \Journal{\PRL}{79}{2204}{1997}.
\bibitem{e787_pnn} Adler~S., {\it et al.}, \Journal{\PRL}{84}{3768}{2000}.
\bibitem{grossman} Grossman~Y., and Nir~Y., \Journal{\PLB}{398}{163}{1997}.
\bibitem{e799_pnn_gg} Adams J, {\it et al.}, \Journal{\PLB}{447}{240}{1999}.
\bibitem{e799_pnn} Alavi-Harati A, {\it et al.}, \Journal{\PRD}{61}{072006}{2000}.
\bibitem{sinb} Bucahalla~G. and Buras~A., \Journal{\PLB}{333}{221}{1994};
Bucahalla~G. and Buras~A., \Journal{\PRD}{54}{6782}{1996};
Nir~Y. and Worah~M.P., \Journal{\PLB}{423}{319}{1998};
Bergmann~S. and Perez~G., hep-ph/0007170.
\end{references}
\end{document}